# Nonhydrostatic Effects and the Determination of Icy Satellites' Moment of Inertia

Peter Gao[1,*] and David J. Stevenson[1]

[1]*Division of Geological and Planetary Sciences, California Institute of Technology, Pasadena, CA, USA, 91125*

*Corresponding author at: *Division of Geological and Planetary Sciences, California Institute of Technology, Pasadena, CA, USA, 91125*

*Email address:* pgao@caltech.edu

*Phone number:* 626-298-9098

**Abstract**

We compare the moment of inertia (MOI) of a simple hydrostatic, two layer body as determined by the Radau-Darwin Approximation (RDA) to its exact hydrostatic MOI calculated to first order in the parameter $q = \Omega^2R^3/GM$, where $\Omega$, R, and M are the spin angular velocity, radius, and mass of the body, and G is the gravitational constant. We show that the RDA is in error by less than 1% for many configurations of core sizes and layer densities congruent with those of solid bodies in the Solar System. We then determine the error in the MOI of icy satellites calculated with the RDA due to nonhydrostatic effects by using a simple model in which the core and outer shell have slight degree 2 distortions away from their expected hydrostatic shapes. Since the hydrostatic shape has an associated stress of order $\rho\Omega^2R^2$ (where $\rho$ is density) it follows that the importance of nonhydrostatic effects scales with the dimensionless number $\sigma/\rho\Omega^2R^2$, where $\sigma$ is the nonhydrostatic stress. This highlights the likely importance of this error for slowly rotating bodies (e.g., Titan and Callisto) and small bodies (e.g., Saturn moons other than Titan). We apply this model to Titan, Callisto, and Enceladus and find that the RDA-derived MOI can be 10% greater than the actual MOI for nonhydrostatic stresses as small as ~0.1 bars at the surface or ~1 bar at the core-mantle boundary, but the actual nonhydrostatic stresses for a given shape change depends on the specifics of the interior model. When we apply this model to Ganymede we find that the stresses necessary to produce the same MOI errors as on Titan, Callisto, and Enceladus are an order of magnitude greater due to its faster rotation, so Ganymede may be the only instance where RDA is reliable. We argue that if satellites can reorient to the lowest energy state then RDA will always give an overestimate of the true MOI. Observations have shown that small nonhydrostatic gravity anomalies exist on Ganymede and Titan, at least at degree 3 and presumably higher. If these anomalies are indicative of the nonhydrostatic anomalies at degree 2 then these imply only a small correction to the MOI, even for Titan, but it is possible that the physical origin of nonhydrostatic degree 2 effects is different from the higher order terms. We conclude that nonhydrostatic effects could be present to an extent that allows Callisto and Titan to be fully differentiated.

## 1. INTRODUCTION

Current models of icy satellite formation and evolution depend on the accuracy with which we determine their interior structures. These can be inferred from their moments of inertia (MOI), which can be estimated from *in situ* gravitational field measurements by spacecraft. The primary method of estimation is the Radau-Darwin approximation (RDA) (e.g., Hubbard 1984, Murray and Dermott 1999), which relates the MOI to the degree 2 response of the body to rotation and tides expressed in the gravitational coefficients $J_2$ and $C_{22}$, defined by:

$$J_2 = \frac{C}{MR^2} - \frac{B+A}{2MR^2} \tag{1a}$$

$$C_{22} = \frac{B-A}{4MR^2} \tag{1b}$$

where M and R are the mass and mean radius of the body, respectively; and C > B > A are the principle moments of inertia of the body. The accuracy of RDA for a two layer body has been explored in works such as Zharkov (2004) and Schubert et al. (2011), where the results of RDA are compared to the exact solution of Kong et al. (2010) and the results of the theory of figures, yielding relatively small differences of ~0.1%. The RDA makes three assumptions: (1) The body is in hydrostatic equilibrium; (2) there are no large density variations; (3) the perturbations arising from tides and rotation are small (i.e., linear response). Our primary focus here is on the first assumption. We note, however, that assumption (2) seems to have been insufficiently explored in the published literature and we accordingly have included brief consideration of this approximation here. Assumption (3) is violated for gas and ice giants because of their rapid rotation and this is the focus of the aforementioned theory of figures discussed in Hubbard (1984) and revisited very recently in Hubbard (2012) in the context of Maclaurin spheroids, the same approach that we use (in the linear limit).

The RDA has been used to determine the MOIs of several large icy satellites, such as Titan (Iess et al. 2010), Ganymede (Anderson et al. 1996), and Callisto (Anderson et al. 2001), as well as medium-sized satellites, such as Rhea (Iess et al. 2007). Enceladus is of great interest but no consensus has emerged yet on the MOI for this body. Our main focus here is on the large icy bodies for which it is commonly assumed that RDA is accurate, but we will also briefly discuss the impact of RDA errors on Enceladus in anticipation of future results. Table 1 shows several physical and orbital parameters for these bodies, including their MOIs determined from RDA if available. Despite the similarities in the masses and radii of Titan, Ganymede, and Callisto, their determined moments of inertia appear to be vastly different: Ganymede's is low, implying full differentiation, while Titan and Callisto's are high, implying partial differentiation. This interpretation has influenced models for satellite formation such as the "gas-starved disk" model for the Galilean moons (Stevenson 2001; Canup and Ward 2002), and alternative ways of reducing accretion heating through lengthening the timescale of accretion (e.g. Mosqueira and Estrada, 2003). These models may avoid differentiation during accretion but differentiation may occur subsequently. In fact, the full differentiation of Ganymede is often attributed to later processes in this model such as tidal heating (Canup and Ward 2002), though differences in the formation environment such as a higher disk temperatures at Ganymede's orbital distance (Barr and Canup 2008) are also proposed. Some models avoid the full differentiation of Callisto and Titan due to later processes by constraining the formation times so as to avoid excessive heating by short-lived radioisotopes (Barr et al. 2010), while others allow for full differentiation of Titan but with a low-density, hydrated silicate core (Fortes 2012; Castillo-Rogez and Lunine 2010). Still others maintain that Callisto and Titan must fully differentiate in the lifetime of the Solar System due to density gradients trapping heat generated by long-lived radioisotopes (O'Rourke and Stevenson 2013).

Ultimately, the usefulness of these models depends on the accuracy of the MOIs that they attempt to explain, which in turn depends on the reliability of the RDA in determining the MOIs from the gravity measurements. The RDA predicts a one-to-one correspondence between MOI and $J_2$ for a specific rotation, but nonhydrostatic effects destroy this correspondence by introducing a nonhydrostatic contribution to $J_2$ that cannot be easily separated out from the measured $J_2$ value. This could be especially troublesome in slowly rotating bodies where the nonhydrostatic contribution could make up a large fraction of the total $J_2$, whereas the effect would be less in fast rotating bodies. For instance, Mueller and McKinnon (1988) noted that nonhydrostatic effects could be present on the slow rotator Callisto in magnitudes that would render its determined MOI untrustworthy. In this paper, we generalize this analysis to any degree 2 nonhydrostatic contribution and extend it to Titan and Ganymede, where long wavelength mass anomalies have been detected (Palguta et al. 2006; Iess et al. 2010). This will allow us to both evaluate the accuracy of Titan and Callisto's determined MOIs given nonhydrostatic effects and determine whether Ganymede is less affected by these effects given its faster rotation. As previously mentioned, we will also extend our analysis to Enceladus due to its unique nature. Considering that a 10% error in the MOI of Callisto and a few % for Titan could potentially result in values consistent with fully differentiated bodies, it is essential that the effects of these nonhydrostatic structures be determined to establish their impact on the calculated MOIs of the large icy satellites, and in turn our understanding of their interiors and evolutionary processes.

In section 2, we first establish the error in RDA assuming exact hydrostatic equilibrium but allowing for large density differences in the context of a nested Maclaurin spheroid model. We then quantify the effects degree 2 nonhydrostatic anomalies have on the MOI of a generalized large icy satellite as determined by the RDA, as well as the relationship between the magnitude of the nonhydrostatic anomaly and the stress caused by such an anomaly on the icy satellite. In section 3 we apply our model to Titan, Ganymede, Callisto, and Enceladus and assess whether existing nonhydrostatic contributions and/or other possible sources are capable of producing major MOI errors. Finally, we summarize our work and state our conclusions in section 4.

## 2. THEORY

### *2.1. Hydrostatic Icy Satellite Model*

The RDA can be expressed as

$$\frac{C}{MR^2} = \frac{2}{3}\left(1 - \frac{2}{5}\sqrt{\frac{5}{3\Lambda_{2,0}+1} - 1}\right) \tag{2}$$

where $C/MR^2$ is the nondimensionalized polar moment of inertia (Hubbard 1984). For a body deformed only by rotation $\Lambda_{2,0} = J_2/q$ in the limit of small values of q, and

$$q = \frac{\Omega^2 R^3}{GM} \tag{3}$$

is a dimensionless measure of the "centrifugal potential" that arise due to rotation, with Ω, R, and M as the spin angular velocity, mean radius, and mass of the body, and G as the gravitational constant. For synchronously rotating satellites, the tidal potential is three times the rotational potential in peak amplitude and this means that $\Lambda_{2,0}$ is replaced by $2J_2/5q$ (or equivalently $\Lambda_{2,0}$ can be replaced in equation 2 by $4C_{22}/3q$ ; note that this implies $J_2/C_{22}=10/3$). But since we are dealing with linear theory it suffices to consider just the rotational disturbance; generalization to include the permanent tide is trivial. In order to simplify calculations and isolate the essential physics of the problem, we will consider a simple rotating two-layer icy satellite model with an ice mantle of constant density $\rho_o$ overlying a rocky core with constant density $\rho_c$. In comparison to real icy satellites, our model ice mantle includes the outer ice shell and any liquid water oceans and higher ice phases that may be present below, while our model rocky core represents the silicate interior. The satellite will be oblate, as it will be distorted by the degree 2 "centrifugal potential". This is therefore a model of two nested Maclaurin spheroids. Large icy satellites have significant variation in ice density within so of course it is not our intent to use this model to obtain accurate interior models. Rather, we are using this simple model to understand the difference between exact results of a simple model and approximate (RDA) results for the same model, with and without nonhydrostatic effects.

In the hydrostatic limit, the two defining surfaces of the model – the surface of the body and the core-mantle boundary – will be slightly oblate due to rotation, with the magnitude of oblateness directly related to the spin rate. We can define these surfaces by

$$r_o = R[1 + \epsilon_o P_2(cos\theta)] \quad (4a)$$

$$r_c = R_c[1 + \epsilon_c P_2(cos\theta)] \quad (4b)$$

where $r_o$ is the radius of the satellite, $r_c$ is the core radius, $R_c$ is the mean core radius, $P_2$ is the degree 2 Legendre polynomial, θ is the colatitude, and the ε's are oblateness factors that give a measure of how flattened the satellite and core surfaces are. The oblateness of the satellite has the effect of introducing a degree 2 term to the external gravitational potential of the satellite, defined by

$$V_2 = \frac{G}{r}\int \rho(\vec{r}')\left(\frac{r'}{r}\right)^2 P_2(cos\theta)P_2(cos\theta')d^3\vec{r}' \quad (5)$$

where r is the distance from the center of the satellite and ρ is the density distribution of the satellite; all primed variables represent positions on or within the satellite. Since the two regions within the body are of constant density and the Legendre functions are orthogonal, the only contributions to the integral come from the deformations of the core and surface. By definition of $J_2$, $V_2$= -$J_2$ $P_2$ $GMR^2/r^3$ whence Equation 5 can be solved to yield

$$J_2 = -\frac{3}{5}\frac{y\epsilon_o + x^5(1-y)\epsilon_c}{y + x^3(1-y)} \quad (6)$$

where $x = R_c/R_o$, $y = \rho_o/\rho_c$, and we have kept terms to only first order in the ε's, as they are of the same order as q and much less than unity for typical icy satellites.

To derive the ε's, we evaluate the satellite and core surfaces as equipotentials. At the satellite surface, the relevant potentials are the degree 0 and 2 gravitational potentials and the "centrifugal potential":

$$V_{surf} = \frac{GM}{r} - \frac{GMR^2}{r^3}J_2P_2 + \frac{1}{3}r^2\Omega^2(1 - P_2) \quad (7a)$$

where we have omitted the colatitudes dependence of $P_2$ for simplicity. We then solve equation 7a on the satellite surface so that

$$V_{surf} = \frac{GM}{R}(1 - \epsilon_O P_2) - \frac{GM}{R}J_2P_2 + \frac{1}{3}R^2\Omega^2(1 - P_2) \quad (7b)$$

where we have retained only terms that are linear in the perturbation (recall that terms such as $\varepsilon J_2$ are second order) . To qualify as an equipotential, the surface must not have any potential variations with latitude or longitude; thus, the summed coefficient of $P_2$ is zero:

$$-\frac{GM}{R}\epsilon_O - \frac{GM}{R}J_2 - \frac{1}{3}R^2\Omega^2 = 0 \quad (7c)$$

. Solving for $\varepsilon_o$ and incorporating equation 3, we get

$$\epsilon_O = -J_2 - \frac{q}{3} \quad (7d)$$

Finally, plugging in equation 6 and simplifying gives:

$$\epsilon_O = \frac{-\frac{5q}{3}[(1-y)x^3+y]+3x^5(1-y)\varepsilon_c}{2y+5x^3(1-y)} \quad (8)$$

Outside the core surface (but beneath the surface), the relevant potentials are the degree 0 and 2 gravitational potentials of the core, the degree 2 gravitational potential of the mantle (which can come only from the surface distortion), and the "centrifugal potential":

$$V_{core} = \frac{GM_c}{r} + \frac{4}{5}\pi G r^2 \rho_o P_2 \epsilon_o + \frac{4}{5}\pi\frac{GR_c^5}{r^3}(\rho_c - \rho_o)P_2\epsilon_c + \frac{1}{3}r^2\Omega^2(1 - P_2) \quad (9)$$

By Gauss' law there is no degree 0 contribution from the mantle. Evaluating at the core surface (equation 4b), we obtain

$$\frac{3}{5}y\epsilon_O = \frac{q_c}{3} + \frac{1}{5}\epsilon_c(2 + 3y) \quad (10)$$

where

$$q_c = q[x^3(1 - y) + y] \quad (11)$$

Combining equations 6, 8, and 10 allows us to solve for $J_2$ and the ε's in terms of q, x, and y. Finally, as we have measured values of the total mass and mean radius of the major icy satellites, we can relate x and y by noting that

$$M = \frac{4}{3}\pi[(\rho_c - \rho_o)R_c^3 + \rho_o R^3] \quad (12a)$$

which can be simplified to

$$y = \frac{x^3}{\tau + x^3 - 1} \quad (12b)$$

where

$$\tau = \frac{3M}{4\pi \rho_o R^3} \quad (12c)$$

is the ratio of mean satellite density to surface density.

The RDA assumes that the bodies in question are devoid of large density variations. To assess this assumption, we use our simple 2-layer model and calculate its exact MOI to first order in q using

$$\frac{C}{MR^2} = \frac{\int s^2 dm}{MR^2} \quad (13a)$$

where s is the distance between a mass element dm of the satellite and the rotation axis. Evaluating equation 13a and simplifying using equation 6, we get

$$\frac{C}{MR^2} = \frac{2}{5}\left[\frac{x^5(1-y)+y}{x^3(1-y)+y}\right] + \frac{2}{3}J_2 \quad (13b)$$

where the first term represents the polar moment of inertia of a spherical body with two constant density layers, and the second term represents the first order correction to the polar moment of inertia due to rotation-induced oblateness. We do not include higher order terms as they are small. We can now compare the exact MOI calculated in equation 13b with the MOI calculated using the RDA. Figure 1 shows the percentage error in the RDA-derived MOI as a function of the core and mantle densities, constrained by Titan's mean density (which determines core radius fraction x) and rotation rate. These models are not attempting to fit the measured MOI but rather to assess the error in RDA for Titan-like models. The results show that the RDA-derived MOIs are <1% different from the exact MOIs for the range of densities of interest. The sign of the error changes in the density range of interest so it is not consistently in one direction. We also see that the absolute errors largely increase with increasing difference between the mantle and core densities, as expected with the assumption in question. Although the error is small, it is noteworthy that there is not an exact one-to one correspondence between the value of $J_2/q$ and $C/MR^2$. This is evident from the fact that the equations determining the true $C/MR^2$ cannot be mapped onto equation 13b. It is indeed surprising that RDA works as well as it does since the exact formulation has so little in common with the formula used in RDA.

Exact results for icy satellites are also given in Zharkov et al. (1985), but they did not examine the differences with RDA in detail. Zharkov (2004) showed that the RDA-derived MOI for Io was within ~0.1% of that derived from the first term of equation 13b, which is in agreement with our results since the second term of equation 13b is small in comparison. They also showed that the inclusion of second order terms in q in the RDA resulted in an order of magnitude decrease in the errors. Meanwhile, results from the comparison of RDA-derived flattening and surface eccentricity in the linear limit to that of the exact solution of Kong et al. (2010) showed errors of comparable magnitude for bodies with internal structures and rotation rates applicable to large icy satellites (Schubert et al. 2011).

To explore all available phase space in density and core size variations, we plot in figure 2 the percent error between the exact and RDA-derived MOIs as a function of the ratio of the core radius to the total radius and the ratio of the density of the outer layer to that of the inner layer for a wider range of values than those appropriate for Titan–like bodies, though we constrain the rotation rate and radius to that of Titan's and the core density to a value of 3 g $cm^{-3}$. Here we see that errors >10% are possible as the density of the outer layer becomes <5% that of the inner layer. The white area indicates parts of the phase space where the absolute error between the exact and RDA-derived MOIs is greater than 50%. Within this region, the error increases rapidly such that, at the bottom of the figure, the error reaches almost 300%. This highlights the limitations of the RDA for bodies with large, extended outer layers surrounding a dense inner layer, such as within the giant planets of both our Solar System and extrasolar planetary systems.

*2.2. Nonhydrostatic Effects and Stress*

In the above analysis, we assumed hydrostatic equilibrium and were thus able to solve for the oblateness factors $\varepsilon_o$ and $\varepsilon_c$ by assuming the satellite and core surfaces were equipotentials. In the case where nonhydrostatic effects are present however, these surfaces would no longer be equipotentials and we then assume that the ε's could be slightly different values. To quantify this effect, we assume that the deviations of the ε's from their hydrostatic values are small, and thus we can Taylor expand the RDA around the hydrostatic value of $\Lambda_{2,0}$, which is directly proportional to $J_2$, which is in turn directly proportional to the ε's. In other words, if we let $C/MR^2 = f(\Lambda_{2,0})$, where f is the right hand side of equation 2, then

$$f\left(\Lambda_{2,0}^{\mathrm{nh}}\right) - f\left(\Lambda_{2,0}^{h}\right) = \left(\frac{df}{d\Lambda_{2,0}}\right)_{\Lambda_{2.0}^{h}} \frac{(a\Delta\epsilon_o + b\Delta\epsilon_c)}{q} \tag{14a}$$

where the nh and h superscripts denote nonhydrostatic and hydrostatic values, respectively; $\Delta\varepsilon_z = \varepsilon_z^{nh} - \varepsilon_z^{h}$, z = o, c; a and b are given by

$$a = -\frac{3}{5}\frac{y}{y + x^3(1-y)} \tag{14b}$$

$$b = -\frac{3}{5}\frac{x^5(1-y)}{y + x^3(1-y)} \tag{14c}$$

and

$$\frac{df}{d\Lambda_{2,0}} = 2\left(\frac{5}{3\Lambda_{2,0}+1} - 1\right)^{-1/2}\left(3\Lambda_{2,0} + 1\right)^{-2} \tag{14d}$$

For the parameters of interest to us, this derivative is ~0.4. Since $C/MR^2$ and $\Lambda_{2,0}$ are both about 1/3 for Titan-like models, this means that a 2.5% difference between $\Lambda^{h}_{2,0}$ and $\Lambda^{nh}_{2,0}$ will lead to a 1% error in $C/MR^2$ . Note that we are treating the nonhydrostatic distortions as independent variables but the reality could be more complicated. For example, if one assumed that the core surface was nonhydrostatic but the outer surface was still hydrostatic then the hydrostatic surface shape would necessarily change (by equation 7d). This is indeed an interesting case since the core (being rock) is more likely to be nonhydrostatic than the shell. In this particular instance, equation 8 tells us that

$$\Delta\varepsilon_o = \frac{3x^5(1-y)\Delta\varepsilon_c}{2y + 5x^3(1-y)} \tag{15a}$$

so that $a\Delta\varepsilon_o + b\Delta\varepsilon_c$ then becomes simply $b^*\Delta\varepsilon_c$ where

$$\frac{b^*}{b} = 5\left[\frac{y+x^3(1-y)}{2y+5x^3(1-y)}\right] \qquad (15b)$$

This exceeds unity but not greatly so for typical parameter values, and is not therefore a major issue in our simple assessment.

Any deviation from hydrostatic equilibrium will have an associated deviatoric stress. This stress could arise in a variety of ways: It could be elastic or it could be viscous or it could be some other more complicated rheological response. In the real body (as distinct from our idealized model) it can arise from density gradients within the material arising from compositional or thermal differences and it may vary spatially in a way that depends on the spatial variability of the rheological parameters (e.g., as in a lithosphere over a viscous substratum, a common model for icy satellites). The precise form of the deviatoric stress field for a realistic model is beyond the intent of this paper. Instead, for our idealized model, we will define the deviatoric stress as the difference between the hydrostatic stress in the presence of the nonhydrostatic structure and the hydrostatic stress in the absence of such a structure, where the hydrostatic stress is defined simply as the weight of the overlying material divided by the area. In other words, the stresses are given by the pressure changes caused by the extra mass associated with the distortion of core or shell:

$$\sigma_o = \rho_o g_o \Delta\epsilon_o R \qquad (16a)$$

$$\sigma_c = (\rho_c - \rho_o) g_c \Delta\epsilon_c R_c \qquad (16b)$$

for the satellite and core surfaces, respectively, where $g_o$ is the gravitational acceleration at the satellite surface calculated by $GM/R^2$, and $g_c$ is the gravitational acceleration at the core surface, given by

$$g_c = \frac{g_o}{x^2(1-y)+y/x} \qquad (17)$$

The actual stress field is of course a tensor so these are not the actual stresses but merely a characteristic scale for these stresses. We note, however, that in the context of viscous relaxation of a homogeneous medium, these values are indeed the order of magnitude of the driving stresses that are responsible for the gradual decay of the nonhydrostatic deformations over long timescales. They are also the order of magnitude for the purely elastic stresses that would result in a uniformly elastic body. They are not the stresses that are present for a body with a more complicated, more realistic structure. For example, a thin elastic shell over a viscous interior can develop hoop stresses in the shell that are larger than the radial stress by as much as the ratio of the radius to the shell thickness. This situation can arise when a body is changing its spin state over billions of years. In those cases the stress that is needed to support the distortion of the shell away from current hydrostatic equilibrium can be much larger than the estimate provided in equation 16a. This would not apply to equation 16b and rocky core deformations, as the core is likely one large mass without any thin elastic outer shells.

Since $\varepsilon \propto q = \Omega^2R^2/gR$ and $\Delta\varepsilon \sim \sigma/\rho gR$, it follows that $\Delta\varepsilon/\varepsilon \sim \sigma/\rho\ \Omega^2R^2$ is the appropriate dimensionless number characterizing the importance of nonhydrostatic effects. Evidently, nonhydrostatic effects are of greatest concern for small or slowly rotating bodies. We can now apply this model to Titan, Ganymede, Ganymede, and Enceladus to evaluate the effect of realistically supportable nonhydrostatic structures on their determined MOIs.

## 3. RESULTS AND DISCUSSION

Figure 3 shows the percent errors in the RDA-derived MOIs of Titan, Ganymede, Callisto and Enceladus caused by degree 2 nonhydrostatic structures related to the amount of stress these structures would apply on these satellites' surfaces and their cores. We see that the nominal stresses corresponding to a 10% error in the determined MOIs for Titan and Callisto are on the order of 0.1 bars for surface stresses and 1 bar for core stresses, while for Ganymede they are on the order of 1 bar for surface stresses and 10 bars for core stresses. For Enceladus, a surface stress of 0.1 bars is also enough to cause a MOI error of 10%, while core stresses of a few bars is necessary to cause the same error. This similarity in behavior arises because Enceladus's faster rotation is compensating for its lower gravity.

These results are consistent with our previous reasoning that faster rotators and larger bodies will require greater nonhydrostatic deviations to produce the same MOI errors as slower rotators and smaller bodies, as a nonhydrostatic structure on a fast rotating/large body would contribute a smaller fraction to the total $J_2$. We have fixed $\rho_o$ and $\rho_c$ to 1 g cm$^{-3}$ and 3 g cm$^{-3}$, respectively, in our results, but they are not much changed by other choices of core and shell density and thickness.

Deviatoric stresses of order bars are perfectly reasonable for a silicate core. By comparison, Earth's Moon is thought to exhibit a fossil tidal bulge that imposes nonhydrostatic stresses of order tens of bars, which has survived for billions of years (Lambeck and Pullan 1980). Similar origins for any deformations within icy satellites cannot be ruled out given their poorly constrained tidal histories. Stresses of order bars are also potentially supportable in the outer regions of the icy shell or throughout the shell by thermal convection, which has been suggested to occur on Titan (Tobie et al. 2006) and Callisto and Ganymede (McKinnon 1997; 2006).

We turn now to the question of whether these nonhydrostatic effects could be permitted by or supported by the data. Two approaches can be made to this question. One is to appeal to observation that the observed $J_2/C_{22}$ is close to 10/3. The other is to appeal to the smallness of higher order harmonics. Both of these arguments are best considered for the Titan data where there have been sufficient flybys to assess the power present in higher order harmonics and precisely determine $J_2/C_{22}$.

We first note that the 10/3 value is the expected result for any rheological model in which the material parameters depend at most on the radial variable. It is thus consistent with (but does not require) a hydrostatic body. However, our main interest here is to assess small deviations away from 10/3. Give the definitions of $J_2$ and $C_{22}$ in equations 1a and 1b, let α be the ratio of the prolate distortion along the satellite-planet direction to the oblate distortion (polar flattening) (B-A)/(C-B). Then

$$\frac{J_2}{C_{22}} = 2 + \frac{4}{\alpha} \tag{18}$$

and the case of a synchronously rotating hydrostatic satellite is α=3. In general, we can write

$$J_2 = J_{2,h} + J_{2,nh} \tag{19a}$$

$$C_{22} = C_{22,h} + C_{22,nh} \tag{19b}$$

where again h means hydrostatic and nh means nonhydrostatic. We can likewise identify the nonhydrostatic and hydrostatic contributions to A, B and C. If the observed gravity field has no other contributions at degree 2 (i.e., $C_{21}$, $S_{21}$, $S_{22}$ are all zero) then the body is at its lowest energy state under the action of tides and rotation. Consider now the case where the body reorients to its lowest energy state by True Polar Wander. This means that the "permanent" rotational and tidal bulges are allowed to be established in accordance with the orientation of the body in space (i.e., they do not follow the reorientation) while the nonhydrostatic part is permitted to rotate into the lowest energy state. This guarantees that $J_{2,nh}$ and $C_{22,nh}$ are positive and are the only non-zero components of the degree two field. Application of RDA to the total measured $J_2$ or $C_{22}$ is then guaranteed to give an overestimate to the MOI. We can attribute an α value (eq. 18) to the nonhydrostatic part alone. If we write $C_{22,nh} = \delta C_{22,h}$ where $0<\delta \ll 1$ then it follows that

$$\frac{J_2}{C_{22}} = \frac{J_{2,h}+J_{2,nh}}{C_{22,h}+C_{22,nh}} = \frac{\frac{10}{3}+\left(2+\frac{4}{\alpha}\right)\delta}{1+\delta} \tag{20a}$$

whence

$$\frac{J_2}{C_{22}} \approx \frac{10}{3}\left[1+\left(\frac{6}{\alpha}-2\right)\frac{\delta}{5}\right] \tag{20b}$$

As expected, this ratio is exactly 10/3 if α = 3. But the deviation from 10/3 is not large even for other values of α. Consider the case of a Titan model where RDA gives $C/MR^2=0.34$ (cf. Iess et al, 2010). Suppose the true value is 0.33, i.e., different by ~3%. From equation 14d and replacing $\Lambda_{2,0}$ by $4C_{22}/3q$, we know that this can arise from a nonhydrostatic correction of 7.5% in $C_{22}$. Setting δ=0.075, we find that $J_2/C_{22}$ deviates from 10/3 by 5(3/α-1)%. We do not know the value of α, but any value larger than 1.5 yields an error of 5% or less. Table 1 of Iess et al. (2010) does not exclude a deviation of 5% (recall that one sigma error bars are quoted). This should not be confused with the formal error in $C_{22}$ which is indeed much smaller. A similar result applies irrespective of how the nonhydrostatic correction is apportioned between $J_2$ and $C_{22}$. Accordingly, the data do not exclude a $C/MR^2$ of 0.33. This value is at the upper end of possible values for a fully differentiated Titan (Castillo-Rogez and Lunine, 2010).

Alternatively, we could estimate nonhydrostatic effects by assuming that the power in higher harmonics is an indication of power in the nonhydrostatic degree 2. From Iess et al. (2010) we see that the degree three harmonics are as large as 5% of degree 2 (but with large uncertainty). This is similar to the example offered above and is therefore consistent with our conclusion about a possible 3% error in $C/MR^2$. This argument is based on a possibly incorrect assumption since the physical source of degree 2 nonhydrostaticity might contribute nothing at degree 3. A possible case of interest is a shape that "froze in" at a time when the satellite was closer to the planet. This would give a false high MOI if RDA were applied. Yet another case, probably of importance for Titan is a nonhydrostatic contribution arising from tidal heating (Nimmo and Bills, 2010). This would also be expected to give positive contributions to the harmonics and therefore yield a higher RDA-derived MOI than the true value.

In the case of Ganymede there were only two flybys and this prevented a reliable determination of higher harmonics (Anderson et al, 2004). However there is no doubt that higher harmonics are needed to describe the data. This led to an analysis based on mass anomalies (Palguta et al, 2006) but insufficient information to perform the analysis we provide here for Titan. In the case of Callisto, there is insufficient data to assess the magnitude of nonhydrostatic effects. Evidently, we require larger nonhydrostaticity for Callisto than is observed for Titan in order to have that body be as differentiated as Titan. It is not even possible to exclude a Ganymede-like MOI for Callisto.

## 4. SUMMARY AND CONCLUSIONS

The primary method for the determination of the internal structures of large icy satellites in our Solar System is the estimation of their moments of inertia (MOI) via the Radau-Darwin Approximation (RDA), which assumes a homogeneous interior in hydrostatic equilibrium. However, these assumptions are challenged by the gravity measurements of Titan, Ganymede, and Callisto, which show these bodies to be at least partially differentiated with clear signals of nonhydrostatic structures. In order to determine the validity of these assumptions, we constructed a simple hydrostatic 2-layer icy satellite model and determined its degree 2 gravity coefficient $J_2$, which is necessary in the use of the RDA. We showed that in the hydrostatic limit the RDA is valid to within 99% for most configurations of core size and internal density variations, including those applicable to all solid bodies in the Solar System. Errors in RDA become >10% when the outer layer becomes <5% as dense as the inner layer, which represents bodies like the gas giants.

To introduce nonhydrostatic effects, we altered the oblateness of the model to nonhydrostatic shapes and calculated the resulting $J_2$'s. This in turn allowed us to relate the change in the RDA-derived MOIs to the changes in the magnitude of the nonhydrostatic effects and thus calculate the errors between the RDA-derived MOIs and the true MOIs of the model due to nonhydrostatic effects. We also calculated the stresses applied to the model by these nonhydrostatic features to evaluate the validity of the MOI errors. We then applied our calculations to Titan, Ganymede, and Callisto. We showed that the RDA-derived MOIs of Titan and Callisto can be ~10% greater than the actual MOIs given nonhydrostatic stresses of ~0.1 bars on the surface or ~1 bar on the core surface; for Ganymede, these two values are an order of magnitude greater. We also applied our calculations to Enceladus, and showed that an MOI error of 10% is possible given nonhydrostatic stresses of ~0.1 bars on the surface and a few bars on the core surface. These results are largely independent of the model parameters used, such as the density of the two layers and the core size. We conclude that the RDA-derived MOIs of slow-rotators and small bodies such as Titan, Callisto, and Enceladus are most affected by nonhydrostatic structures, while faster rotators and/or larger bodies like Ganymede are less affected.

This is especially important for future analysis of Enceladus gravity measurements in determining its MOI given the large nonhydrostatic contribution from its southern polar depression.

The nonhydrostatic structures corresponding to the aforementioned stresses have not been ruled out in the case of Titan given recent gravity measurements from Cassini, while data for Ganymede, Callisto, and Enceladus are insufficient to provide any bounds on the degree of nonhydrostatic structures present. Therefore, the difference in the observed MOIs of Titan/Callisto and Ganymede – and thus the perceived dichotomy of their internal structures – may be entirely due to nonhydrostatic effects, and that it is possible all three bodies are fully differentiated.

Our work shows that caution must be exercised in determining the MOIs of the outer Solar System icy satellites using the RDA, and that the nonhydrostatic contributions to their gravity fields must be better characterized if we are to rely on their MOIs to construct models of their interior structures, their formation, and their evolution.

## REFERENCES

Anderson J. D., Lau E. L., Sjogren W. L., Schubert G., and Moore W. B. (1996) Gravitational constraints on the interior structure of Ganymede. *Nature*, 384, pp. 541-543.

Anderson J. D., Jacobson R. A., McElrath T. P., Moore W. B., Schubert G., Thomas P. C. (2001) Shape, mean radius, gravity field, and internal structure of Callisto. *Icarus*, 153, pp. 157-161.

Anderson, J.D., Schubert, G., Jacobson, R.A., Lau, E.L., Moore, W.B., Palguta, J., 2004. Discovery of mass anomalies on Ganymede. Science 305, 989–991.

Barr A. C. and Canup R. M. (2008) Constraints on gas giant satellite formation from the interior states of partially differentiated satellites. *Icarus*, 198, pp. 163-177.

Barr A. C., Citron R. I., and Canup R. M. (2010) Origin of a partially differentiated Titan. *Icarus*, 209, pp. 858-862.

Canup R. M. and Ward W. R. (2002) Formation of the Galilean satellites: conditions of accretion. *Astron. J.*, 124, pp. 3404-3424

Castillo-Rogez J. C. and Lunine J. I. (2010) Evolution of Titan's rocky core constrained by Cassini observations. *Geophys. Res. Lett.*, 37, L20205, 5pp.

Fortes A. D. (2012) Titan's internal structure and the evolutionary consequences. *Planet. Space. Sci.*, 60, pp. 10-17.

Hubbard W. B. (1984) In: *Planetary Interiors*. Von Nostrand Reinhold Company Inc., New York, New York, USA.

Hubbard W. B. (2012) High-precision Maclaurin-based models of rotating liquid planets. *Astrophys. J. Lett.*, 756, L15, 3pp.

Iess L., Rappaport N. J., Tortora P., Lunine J., Armstrong J. W., Asmar S. W., Somenzi L., and Zingoni F. (2007) Gravity field and interior of Rhea from Cassini data analysis. *Icarus*, 190, pp. 585-593.

Iess L., Rappaport N. J., Jacobson R. A., Racioppa P., Stevenson D. J., Tortora P., Armstrong J. W., and Asmar S. W. (2010) Gravity field, shape, and moment of inertia of Titan. *Science*, 327, pp. 1367-1369.

Jacobson R. A., Antreasian P. G., Bordi J. J., Criddle K. E., Ionasescu R., Jones J. B., Mackenzie R. A., Meek M. C., Parcher D., Pelletier F. J., Owen W. M. Jr., Roth D. C., Roundhill I. M., and Stauch J. R. (2006) The gravity field of the Saturnian system from satellite observations and spacecraft tracking data. *Astron. J.*, 132, pp. 2520-2526.

Kong D., Zhang K., and Schubert G. (2010) Shapes of two-layer models of rotating planets. *J. Geophys. Res.*, 115, E12003, 11pp.

Lambeck K. and Pullan S. (1980) The lunar fossil bulge hypothesis revisited. *Phys. Earth Planet. In.*, 22, pp. 29-35.

McKinnon W. B. (1997) Mystery of Callisto: Is it undifferentiated? *Icarus*, 130, pp. 540-543.

McKinnon W. B. (2006) On convection in ice I shells of outer Solar System bodies, with detailed application to Callisto. *Icarus*, 183, pp. 435-450.

Mosqueira I. and Estrada P. R. (2003) Formation of the regular satellites of giant planets in an extended gaseous nebula I: subnebula model and accretion of satellites. *Icarus*, 163, pp. 198-231.

Mueller S. and McKinnon W. B. (1988) Three-layered models of Ganymede and Callisto: compositions, structures, and aspects of evolution. *Icarus*, 76, pp. 437-464.

Murray C. D. and Dermott S. F. (1999) In: *Solar System Dynamics*. Cambridge University Press, New York, New York, USA.

Nimmo F. and Bills B. G. (2010) Shell thickness variations and the long-wave topography of Titan. *Icarus*, 208, pp. 896-904.

O'Rourke J. and Stevenson D. J. (2013) Stability of ice/rock mixtures with application to a partially differentiated Titan. *Icarus*, in revision.

Palguta J., Anderson J. D., Schubert G., and Moore W. B. (2006) Mass anomalies on Ganymede. *Icarus*, 180, pp. 428-441.

Schubert G., Anderson J., Zhang K., Kong D., and Helled R. (2011) Shapes and gravitational fields of rotating two-layer Maclaurin ellipsoids: Application to planets and satellites. *Phys. Earth Planet. In.*, 187, pp. 364-379.

Showman A. P. and Malhotra R. (1999) The Galilean satellites. *Science*, 286, pp. 77-84.

Stevenson D. J. (2001) Jupiter and its moons. *Science*, 294, pp. 71-72.

Tobie G., Lunine J. I., and Sotin C. (2006) Episodic outgassing as the origin of atmospheric methane on Titan. *Nature*, 440, pp. 61-64.

Zharkov V., Leontjev, V.V. and Kozenko, A. V. (1985) Models, Figures and Gravitational Moments of the Galilean satellites of Jupiter and Icy Satellites of Saturn. *Icarus* 61, 92-100.

Zharkov V. N. (2004) A theory of the equilibrium figure and gravitational field of the Galilean satellite Io: The second approximation. *Astron. Lett+.*, 30, pp. 496-507.

| | Titan | Ganymede | Callisto | Enceladus |
|---|---|---|---|---|
| **Mass ($10^{26}$ g)** | 1.3452 ± 0.0002[c] | 1.48167 ± 0.0002[d] | 1.0759 ± 0.0001[e] | 0.00108 ± 1e-6[c] |
| **Mean Radius (km)** | 2574.73 ± 0.09[f] | 2634.1 ± 0.3[g] | 2408.4 ± 0.3[g] | 252.1 ± 0.1[c] |
| **Orbital Period (Earth days)[a]** | 15.95[h] | 7.15[h] | 16.69[h] | 1.37[h] |
| **MOI[b]** | 0.3414 ± 0.0005[f] | 0.3105 ± 0.0028[d] | 0.3549 ± 0.0042[e] | Unknown |
| **q ($10^{-5}$)[i]** | 3.9545 | 19.131 | 3.6958 | 626.374 |

[a] Assumed to be the same as rotation period, i.e. synchronous rotation.
[b] Moment of inertia in units of $C/MR^2$, where C, M, and R are the polar moment of inertia, mass, and mean radius of the body in question.
[c] Jacobson et al. 2006.
[d] Anderson et al. 1996.
[e] Anderson et al. 2001. Callisto mass and associated uncertainty calculated from dividing given GM value and GM uncertainty by given G value.
[f] Iess et al. 2010 (SOL1 flybys)
[g] Showman and Malhotra 1999.
[h] Murray and Dermott 1999 (no uncertainties were given).
[i] $q = \Omega^2 R^3/GM$, where Ω, R, and M are the spin/orbital angular velocity, radius, and mass, respectively, of the satellite; and G is the gravitational constant.

Table 1. Selected physical and orbital parameters of Titan, Ganymede, Callisto, and Enceladus.

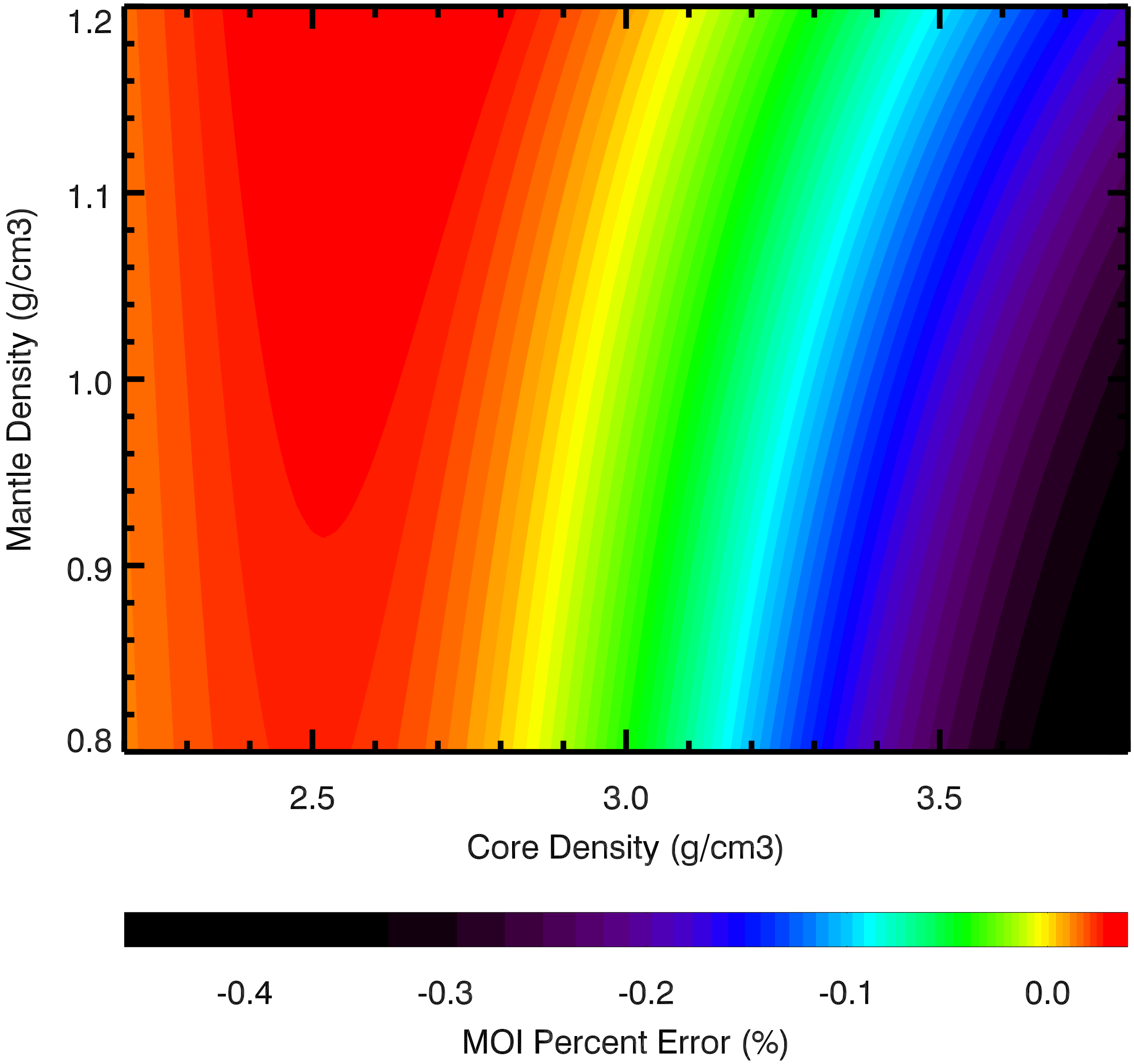

Figure 1. The percent error between the true moment of inertia of a hydrostatic 2-layer icy satellite calculated to first order in q and one calculated by the Radau Darwin Approximation as a function of the mantle and core densities. The model is constrained by the density and rotation rate of Titan.

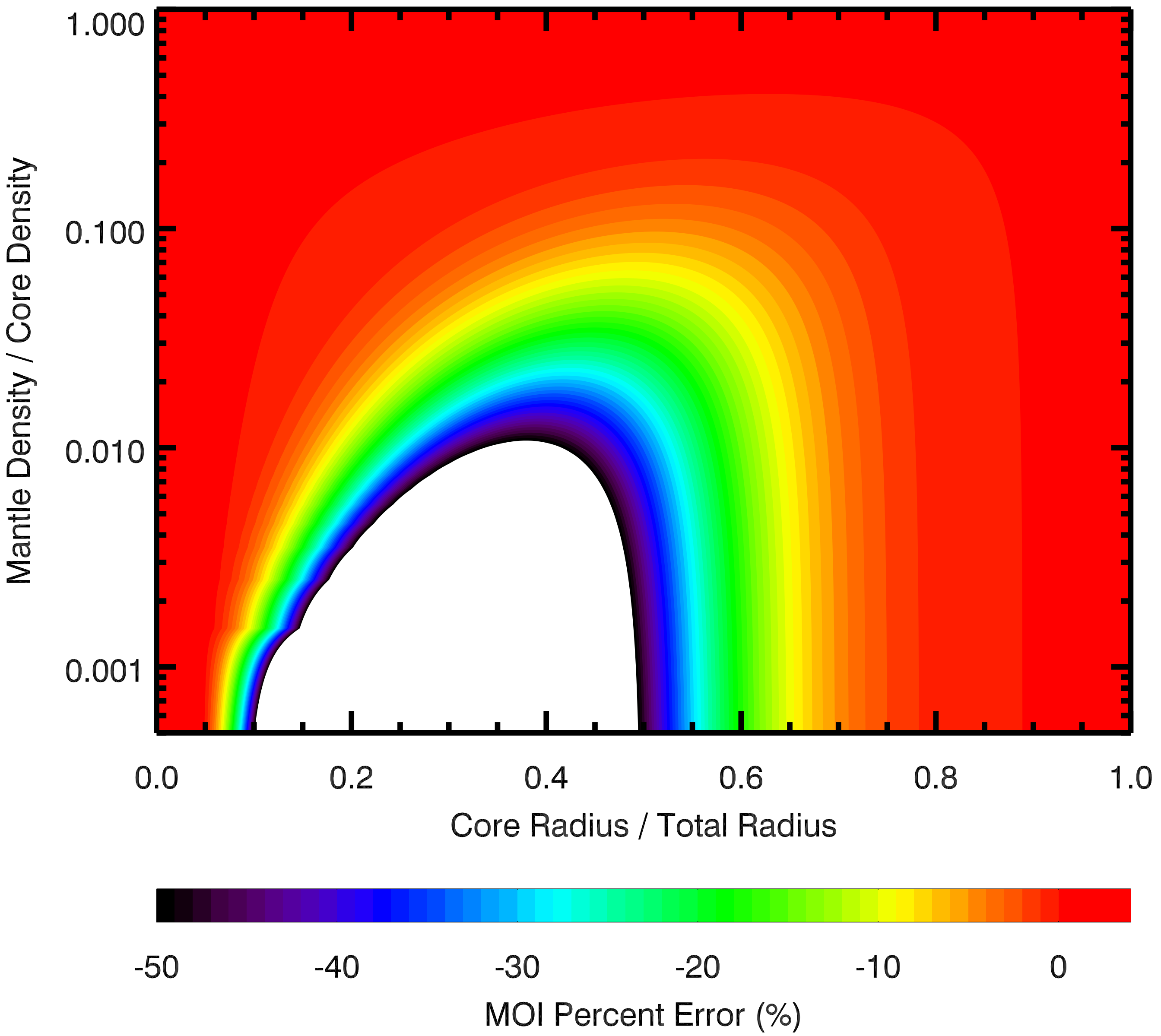

Figure 2. The percent error between the true moment of inertia of a hydrostatic 2-layer oblate body calculated to first order in q and one calculated by the Radau Darwin Approximation as a function of the core radius to total radius ratio, and the outer layer density to inner layer density ratio. The model is constrained by the radius and rotation rate of Titan and a density of 3 g $cm^{-3}$ for the core. The white area in the lower left represents all percent error values below -50%, with a maximum absolute error within the range of the plot of ~293%. The y axis is in log scale to emphasize the increase in percent error as the ratio of the layers' densities approaches zero.

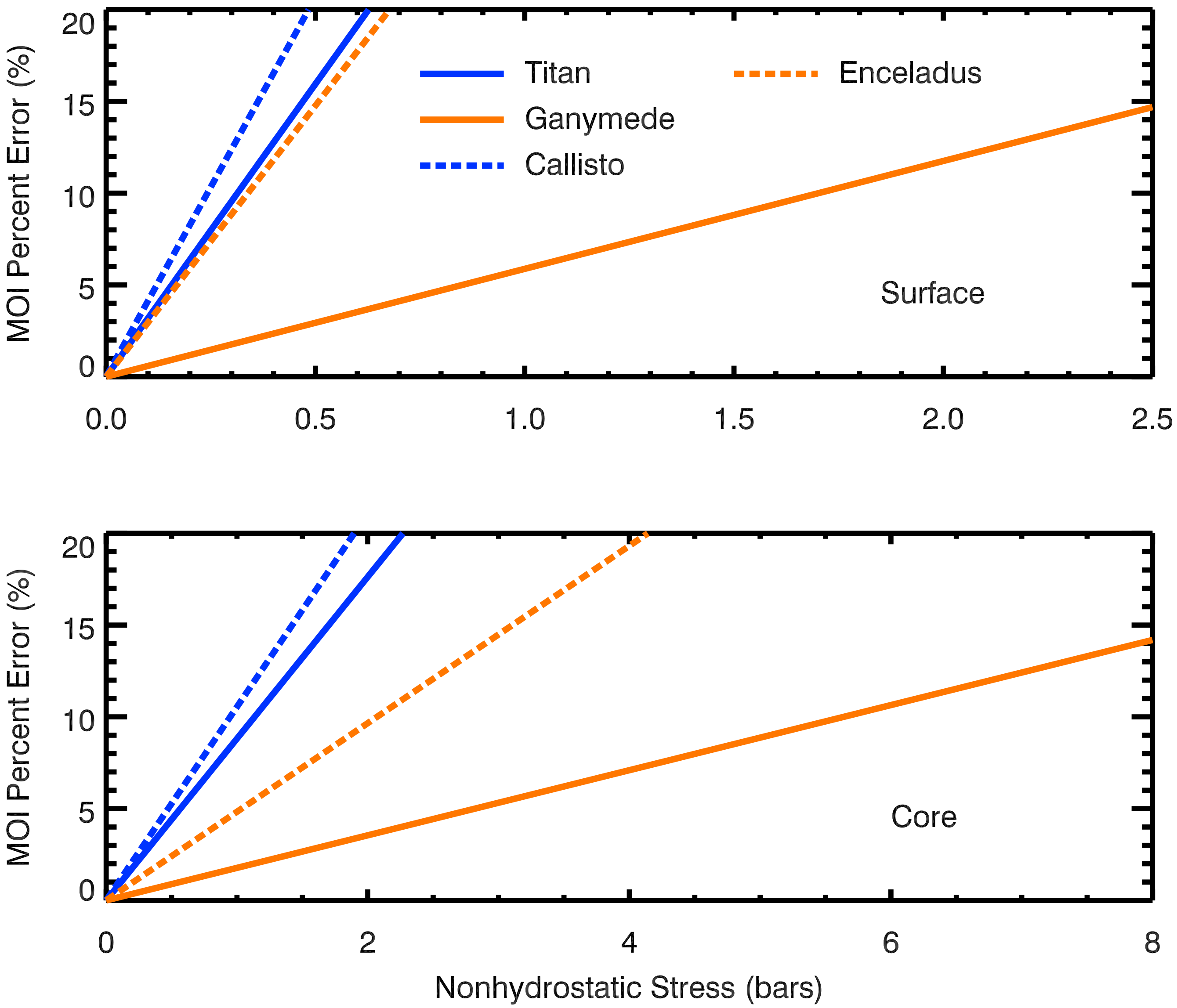

Figure 3. The percent error in the moment of inertia calculated by the Radau Darwin Approximation for Titan, Ganymede, Callisto, and Enceladus due to a degree 2 nonhydrostatic structure on the satellite surface (top) and core mantle boundary (bottom), as a function of the stress imparted by such a structure on the satellite.